\documentclass{emulateapj}

\usepackage{graphicx}

\shorttitle{Growth of Magnetic Field}
\shortauthors{Cho \&  Yoo}

\begin{document}
\title{Growth of a localized seed magnetic field in a turbulent medium} 
\author{Jungyeon Cho and Hyunju Yoo}
\affil{Dept. of Astronomy and Space Science, Chungnam National University, Daejeon, South Korea; jcho@cnu.ac.kr}

\begin{abstract}
Turbulence dynamo deals with amplification of a seed magnetic field in a turbulent medium
and has been studied mostly for uniform or spatially homogeneous seed magnetic fields.
However, some astrophysical processes (e.g. jets from active galaxies, galactic winds, 
or ram-pressure stripping in galaxy clusters)
can provide localized seed magnetic fields.
In this paper, we numerically study amplification of localized seed magnetic fields in a turbulent medium.
Throughout the paper, we assume that driving scale of turbulence is comparable to the size of the system.
Our findings are as follows.
First, turbulence can amplify a localized seed magnetic field very efficiently.
The growth rate of magnetic energy density is as high as that for a uniform seed magnetic field.
This result implies that a magnetic field ejected from an astrophysical object can be a 
viable source of 
magnetic field in a cluster.
Second, the localized  seed magnetic field disperses and fills the whole system very fast.
If turbulence in a system (e.g.~a galaxy cluster or a filament) is driven at large scales, 
 we expect that it takes a few large-eddy turnover times
for magnetic field to fill the whole system.
Third, 
growth and turbulence diffusion of a localized seed magnetic field are also
fast in high magnetic Prandtl number turbulence.
Fourth,  
even in 
 decaying turbulence, a localized seed magnetic field can ultimately fill the whole system.
 Although the dispersal rate of magnetic field is not fast in purely decaying turbulence,
  it can be enhanced by an additional forcing.

\end{abstract}
\keywords{ISM:general - intergalactic medium - MHD - turbulence}       
 
\section{Introduction}

The origin and growth of magnetic field in the universe is an important unsolved problem
in astrophysics (Kulsrud \& Zweibel 2008).
Earlier studies have shown that turbulence can efficiently amplify a weak seed magnetic field.
The main mechanism that is responsible for the amplification is stretching of magnetic field lines 
(Batchelor 1950; Zel'dovich et al. 1984; Childress \& Gilbert 1995).
This process is called \textit{turbulence dynamo} (or \textit{small-scale dynamo}) and 
should not be
confused with \textit{mean field dynamo} (or \textit{large-scale dynamo}), which deals with
generation/growth of a large-scale  magnetic field. 

Small-scale turbulence dynamo has been studied since 1950s (see,
e.g., Batchelor 1950;  Kazantsev 1968; Kulsrud \& Anderson 1992; Kulsrud et al.~1997; 
Cho \& Vishniac 2000; Schekochihin et al. 2004; Brandenburg \& Subramanian 2005; 
Schekochihin \& Cowley 2007; Ryu et al. 2008; Cho et al. 2009). 
When we introduce a weak \textit{uniform} magnetic field in a turbulent medium, 
amplification of the uniform field happens in three stages 
(see Schekochihin \& Cowley 2007; see also Cho \& Vishniac 2000, Cho et al.~2009). 
(1) Stretching of
magnetic field lines occurs most actively near the velocity dissipation
scale first, and the magnetic energy grows exponentially. 
Note that eddy motions in turbulence are fastest at the velocity dissipation scale.
(2) The exponential growth stage ends when the magnetic energy
becomes comparable to the kinetic energy at the dissipation
scale. The subsequent stage is characterized by a linear growth
of magnetic energy and a gradual shift of the peak of magnetic
field spectrum to larger scales. 
(3) The amplification of magnetic
field stops when the magnetic energy density becomes comparable
to the kinetic energy density and a final, statistically steady, saturation stage
begins.

Knowledge on turbulence dynamo is useful for understanding the origin of magnetic fields 
in many astrophysical fluids.
For example, Ryu et al. (2008) numerically studied 
amplification of the intergalactic magnetic field
 by the dynamo action of turbulence.
They assumed that turbulence is generated by cosmological shocks, 
which are produced when primordial gases fall into the gravitational potential well 
caused by dark matters
in clusters and filaments (Ryu et al. 2003).
For simplicity, they assumed that a weak uniform seed magnetic field fills the whole system 
at the beginning of the simulation.
They found that turbulence is strong in the intracluster medium (ICM) and, therefore, 
has almost reached
the saturation stage and that turbulence in filaments is weak and in early linear stage.
Based on a more quantitative analysis, they estimated that the
average magnetic field strength would be a few $\mu$G inside clusters/groups,
approximately 0.1 $\mu$G around clusters/groups, and approximately 10 nG in
filaments.

Existence of regular magnetic fields in many astrophysical fluids, 
such as the interstellar medium (ISM), 
justifies the use of uniform seed magnetic fields in turbulence dynamo studies.
However, there are astrophysical systems where existence of regular fields is uncertain.
For example, it is not clear whether or not regular fields exist in the ICM.
If seed magnetic fields are relic primordial fields  
(Rees 1987; Kronberg 1994; 
Dolag, Bartelmann \& Lesch 1999, 2002;  Gnedin, Ferrara \& Zweibel 2000; 
Banerjee \& Jedamzik 2003; Murgia et al. 2004;
Br\"uggen et al. 2005; Dubois \& Teyssier 2008),
 the correlation length of the seed fields 
can be very large and
we may have regular fields in the ICM.
On the other hand, if seed magnetic fields have astrophysical origins 
(e.g. jets from active galaxies,  galaxy winds, and ram-pressure stripping, etc.;
 see Rephaeli 1988 for an earlier discussion on the origin of the ICM magnetic fields 
 through ram-pressure stripping),
we will probably have no regular fields in the ICM.
In the latter case, we expect that seed magnetic fields ejected from astrophysical 
objects are highly localized
in space.
Then a question arises: how do these localized seed magnetic fields grow?
In this paper, we study amplification of localized seed magnetic fields in a turbulent medium.

Earlier cosmological simulations have shown that  magnetic fields ejected 
{}from active galactic nuclei (AGNs)  
(Ensslin et al. 1997;  V\"olk \& Atoyan 2000; Donnert et al 2009; Xu et al. 2010, 2011) 
or normal galaxies
through
galactic winds (Kronberg, Lesch \& Hopp 1999; Bertone, Vogt \& Ensslin 2006)
or ram-pressure stripping (Arieli, Rephaeli \& Norman 2011)
can be  dispersed and amplified by turbulence and are viable origins of magnetic fields 
in galaxy clusters.
Xu et al. (2010), for example, 
showed that ``as long as the AGN magnetic fields are ejected before the major mergers 
in the cluster formation history, magnetic
fields can be transported throughout the cluster and can be further amplified by the ICM
turbulence caused by hierarchical mergers during the cluster formation process.'' 
However, those simulations are compressible ones and it is usually difficult 
to obtain converging results free from numerical effects in compressible turbulence simulations
(see discussions in Ryu et al.~2008).
Therefore it is advantageous to use an incompressible code because we can control numerical viscosity
and resistivity in incompressible codes. 

In this paper, we use an incompressible code with hyper-viscosity and hyper-diffusion, so that 
we have virtually zero
intrinsic numerical viscosity and diffusion, and the maximized inertial range.
We assume that driving scale of turbulence is comparable to the size of the whole system,
which would be true if turbulence is produced by cosmological shocks or major mergers 
in galaxy clusters.
We also assume that the seed magnetic field has a doughnut shape at t=0.
We describe numerical method in Section 2, and we present results in Section 3.
We give discussions in Section 4 and the summary in Section 5.

\section{Numerical Methods}
We use a pseudospectral code to solve the 
incompressible MHD equations in a periodic box of size $2\pi$:
\begin{equation}
\frac{\partial {\bf v} }{\partial t} = -(\nabla \times {\bf v}) \times {\bf v}
      +(\nabla \times {\bf B})
        \times {\bf B} + \nu \nabla^{2} {\bf v} + {\bf f} + \nabla P' ,
        \label{veq}
\end{equation}
\begin{equation}
\frac{\partial {\bf B}}{\partial t}= 
     \nabla \times ({\bf v} \times{\bf B}) + \eta \nabla^{2} {\bf B} ,
     \label{beq}
\end{equation}
\begin{equation}
      \nabla \cdot {\bf v} =\nabla \cdot {\bf B}= 0,
\end{equation}
where $\bf{f}$ is random driving force,
$P'\equiv P + {\bf v}\cdot {\bf v}/2$, ${\bf v}$ is the velocity,
and ${\bf B}$ is magnetic field divided by $(4\pi \rho)^{1/2}$.
We use 22 forcing components with $2\leq k \leq \sqrt{12}$.
Each forcing component has correlation time of $\sim 1$.
The peak of energy injection occurs at $k\approx 2.5 $.
In our simulations with $\nu=\eta$, 
$v\sim 1$ during the growth stage and $\sim 0.8$ during the saturation stage (see Figure 1).
Therefore, one eddy turnover time at the outer scale of turbulence (or large-eddy turnover time), 
$\sim L/v$, is approximately $\sim$2.5 and $\sim$3 time units, respectively.
Since $v\sim 1$, ${\bf v}$ can be viewed as roughly the velocity 
measured in units of the r.m.s. velocity
of the system and ${\bf B}$ as the Alfv\'en speed in the same units.
Other variables have their usual meaning.

In pseudospectral methods, we calculate the temporal evolution of
the equations (\ref{veq}) and (\ref{beq}) in Fourier space.
To obtain the Fourier components of nonlinear terms, we first calculate
them in real space, and transform back into Fourier space.
We use exactly same forcing terms
for all simulations.
We use an appropriate projection operator to calculate 
$\nabla P'$ term in
{}Fourier space and also to enforce divergence-free condition
($\nabla \cdot {\bf v} =\nabla \cdot {\bf B}= 0$).
We use up to $512^3$ collocation points.

At $t=0$, 
the magnetic field has a doughnut shape, which mimics a magnetic field ejected by  
a galactic outflow\footnote{
  Note however that we use a weak uniform seed magnetic field for the reference runs REF256 and REF256-Pr, 
  in which
  the strength of the mean field, $B_0$, is 0.0001 and we have only the mean field at t=0.
   }.
We use the following expression for the magnetic field at t=0:
\begin{equation}
  {\bf B}(r_{\bot},\Delta x) 
  = \frac{ B_{max}}{ 2 \sigma_0^2 e^{-1}}  r_{\bot} ^2 e^{ -r_{\bot}^2 /2 \sigma_0^2 } 
                             e^{ -\Delta x^2/8 \sigma_0^2 } 
     \hat{\bf \theta}_{\perp},   \label{eq:b_shape}
\end{equation}
where $\sigma_0=4\sqrt{2}$, $r_{\bot}=( \Delta y^2+ \Delta z ^2 )^{1/2}$, 
and $\Delta x, \Delta y,$ and $\Delta z$ are distances measured from the center of the
numerical box in grid units. The unit vector $ \hat{\theta}_{\perp}$ is perpendicular to
$(\Delta x, 0,0)$ and $(0,\Delta y, \Delta z)$.
Note that the maximum strength of the magnetic field at t=0 is $B_{max}$.
Since $\sigma_0=4\sqrt{2}$ in Equation (\ref{eq:b_shape}), the size of
the magnetized region at t=0 is 
$\sim 16$ in grid units, which is
$\sim 1/16$ of the whole system for runs with $256^3$ grid points and
$\sim 1/32$ for $512^3$.
Therefore, in a cluster of size$\sim 1Mpc$, the size of the initially magnetized region corresponds to 
$\sim 60kpc$ and $\sim 30kpc$, respectively.

Hyperviscosity and hyperdiffusion are used for dissipation terms.
The power of hyperviscosity
is set to 3, such that the dissipation term in the above equation
is replaced with
\begin{equation}
 -\nu_n (\nabla^2)^n {\bf v},
\end{equation}
where $n=$ 3.
The same expression is used for the magnetic dissipation term.
We list parameters used for the simulations in Table \ref{table_1}.
Diagnostics of our code can be found in Cho \& Vishniac (2000).

\begin{figure*}
\includegraphics[scale=0.65,bb=0 0 380 400]{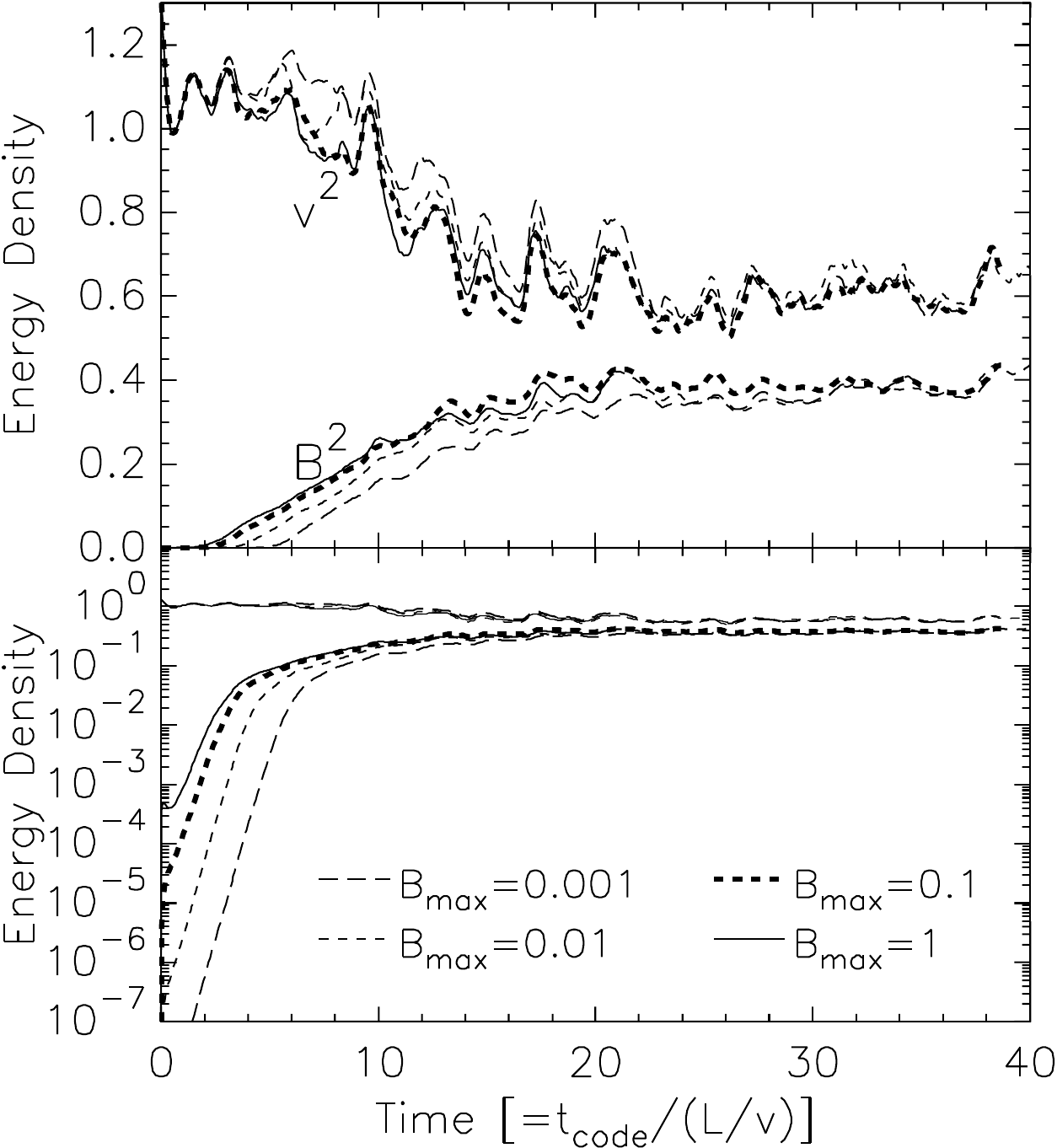}
\hfill
\includegraphics[scale=0.65,bb=0 0 380 400]{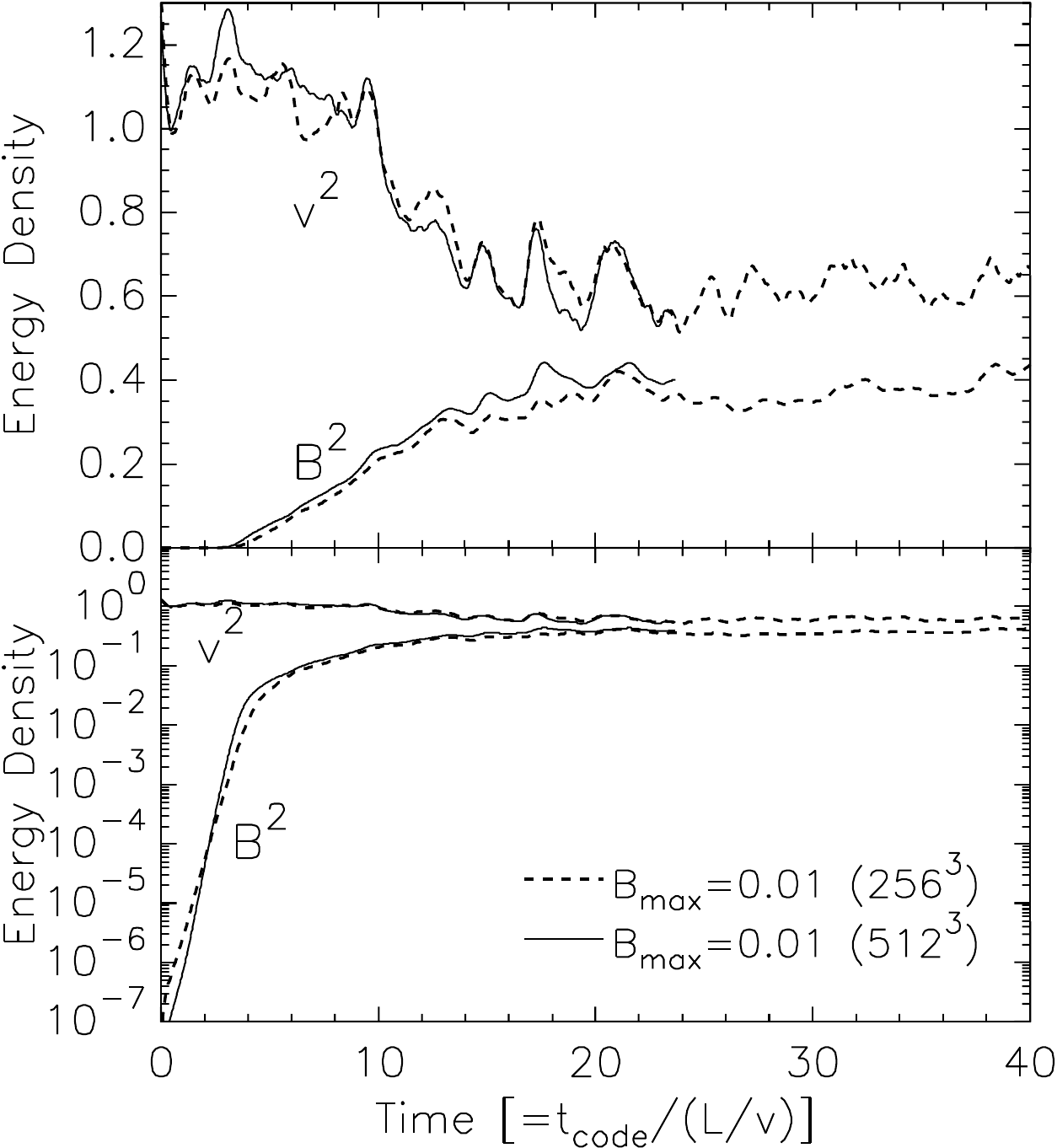}
\caption{Time evolution of $B^2$ and $v^2$ in runs with $\nu=\eta$. 
     Each run starts from a fully developed turbulent velocity field and
      a localized seed magnetic field.
     The localized seed magnetic fields have a doughnut shape.
     All runs have the same numerical resolution.
     Different lines denote different values of $B_{max}$.
     Time is given in units of the large-eddy turnover time defined by $L/v$: $t \equiv t_{code}/(L/v)$, where
     $t_{code}$ is time in code units, $L$ ($\sim$ 2.5) is the outer-scale of turbulence and
     $v$ ($\sim$ 1) is the r.m.s. velocity \textit{before} the saturation stage.
     Results are from 256-$B_{max}$0.001, 256-$B_{max}$0.01, 256-$B_{max}$0.1, and 256-$B_{max}$1.
    Note the logarithmic scale for the vertical axis in the lower panel.
 }
\label{fig:en_many}
\caption{Resolution study. We compare results of 256-$B_{max}$0.01 and 512-$B_{max}$0.01. 
      Two runs have different numerical resolutions, but other numerical setups are the same.
      Two results show good agreements.
       Note the logarithmic scale for the vertical axis in the lower panel.
}
\label{fig:comp_512}
\end{figure*}

\begin{figure*}
\includegraphics[width=0.49\textwidth,bb=0 0 500 330]{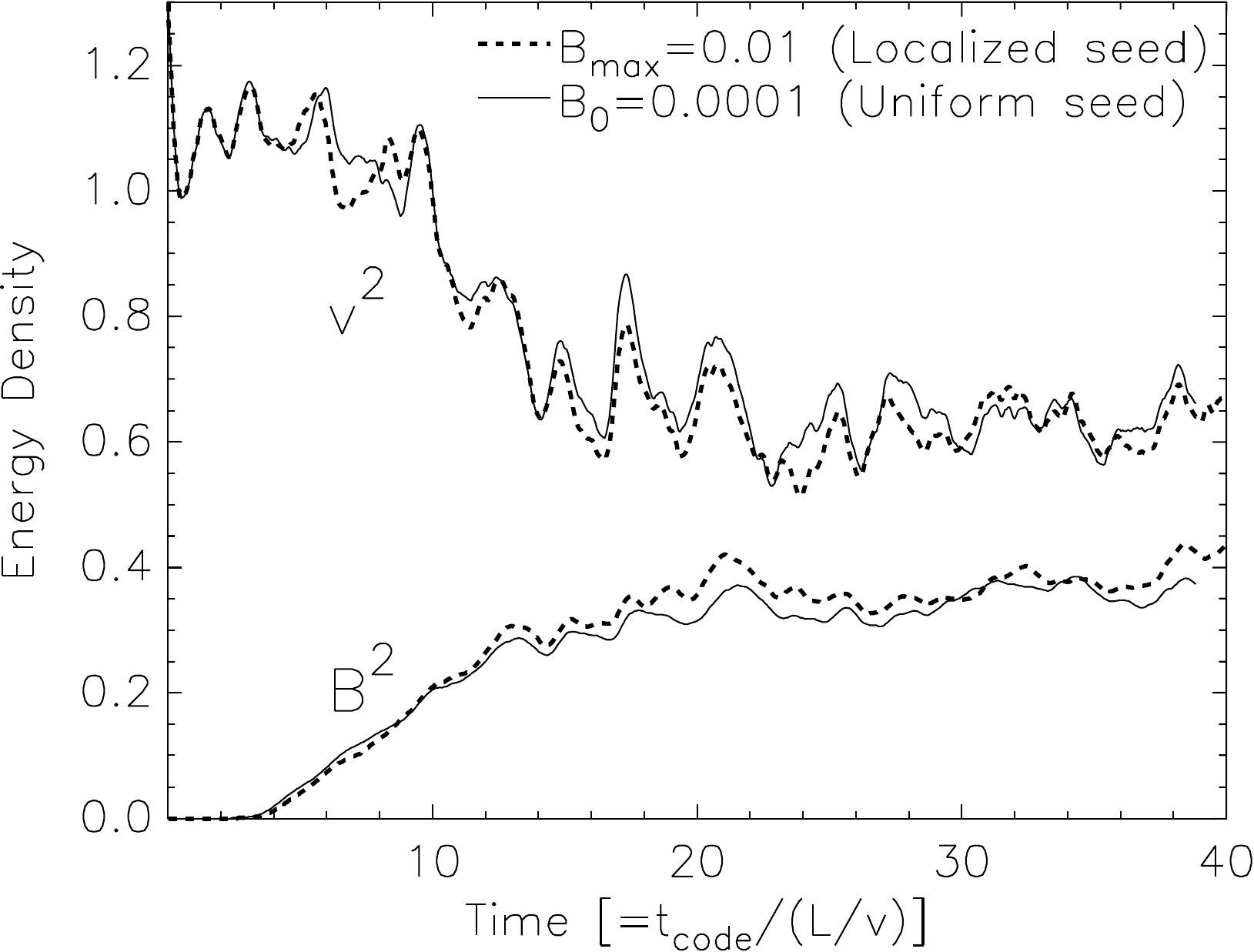}
\includegraphics[width=0.49\textwidth,bb=0 0 500 330]{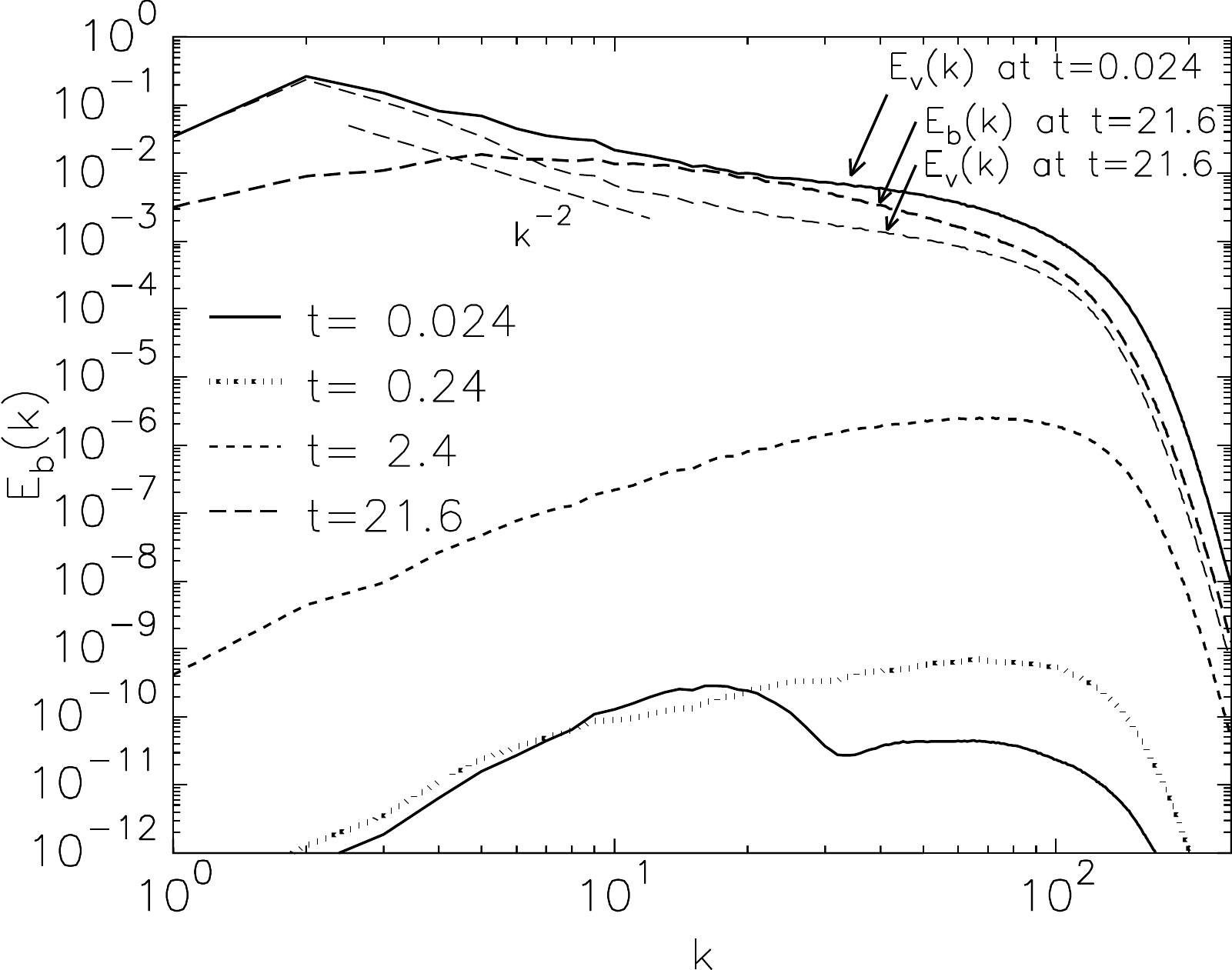}
\caption{Comparison with the case of a uniform seed magnetic field. 
       We compare the results of 256-$B_{max}$0.01 and REF256.
      Note that we use a localized seed magnetic field in the former and a uniform seed field ($B_0=0.0001$)
     in the latter. We find a good correspondence between two results.
}
\label{fig:comp_unif}
\caption{Time evolution of magnetic spectrum.
     The initial seed magnetic field (see the solid curve) looses its original shape quickly and forms a 
    spectrum that has a peak near the dissipation scale (see the dotted curve).
   Then the spectrum goes up without changing the shape much (see the dashed line).
   At this point, the seed magnetic field already fills the most of the simulation box 
   (see Figure \ref{fig:diffusion}
  and Figure \ref{fig:sig}). After magnetic field  fills the whole volume, time evolution of 
  magnetic spectrum and
  energy density should be very similar to those of weak uniform seed magnetic field cases. 
  Time is given in units of $L/v$.
   {}From Run 512-$B_{max}$0.01.
   }
\label{fig:sp}

\end{figure*}

\section{Results}
\subsection{Turbulence with unit magnetic Prandtl number ($\nu=\eta$)}
Figure \ref{fig:en_many} shows time evolution of $v^2$ and $B^2$ for different values of $B_{max}$.
The initial magnetic field has a doughnut shape (see Equation [\ref{eq:b_shape}]) and $B_{max}$ is the maximum
strength of the initial magnetic field.
As in the uniform seed magnetic field cases, growth of magnetic energy density occurs in three stages.
Magnetic energy grows exponentially first.
We can see the exponential growth in the lower panel of Figure \ref{fig:en_many}, 
where we use a logarithmic scale for the y-axis.
When $B_{max}$ is small (see, for example, 256-$B_{max}$0.001 and 256-$B_{max}$0.01),
the growth rate is virtually independent of the value of $B_{max}$.
Then a slower linear growth stage follows, which we can clearly observe in the upper panel of 
the figure.
All runs show similar growth rates in this stage.
Finally, after $t\sim 20$ in units of $L/v$ ($\sim$ 2.5 in code time units), 
a statistically stationary saturation stage is reached.
The saturation level of magnetic energy density is comparable to the kinetic energy density.
Figure \ref{fig:en_many} shows that the stronger the seed magnetic field (or $B_{max}$) is, the earlier
the linear growth stage begins and the sooner the saturation stage is reached.

Resolution study shows that
the linear growth rate does not strongly depend on the numerical resolution.
In Figure \ref{fig:comp_512}, we compare Runs 512-$B_{max}$0.01 and 256-$B_{max}$0.01, in which
the shape of seed magnetic field (see Equation [\ref{eq:b_shape}]) and 
numerical setup are identical and only the numerical resolutions are different.
As we can see in the figure, the overall results are very similar. 
That is, the growth rates at the linear growth stage and 
the saturation levels of the magnetic energy density are very similar.
In the lower panel of the figure, we plot the same quantities, but we use a logarithmic scale for the y-axis.
The lower panel shows the growth rate of Run 512-B$_{max}$0.01 during the exponential growth stage 
is higher than that of 256-B$_{max}$0.01, 
which is due to the fact that the eddy turnover time at the velocity
dissipation scale of Run 512-B$_{max}$0.01 is shorter.

According to Figure \ref{fig:en_many}, the growth rate of magnetic energy during the linear growth stage 
seems to be independent of the value of $B_{max}$.
The linear growth rate in Figure \ref{fig:en_many} is also in good correspondence with 
that of uniform seed magnetic field cases.  
When we compare the linear grow rates of Run 256-B$_{max}$0.01 (localized seed magnetic field) 
and Run REF256 (uniform seed magnetic field), they are very similar (Figure \ref{fig:comp_unif}).
The saturation levels of the magnetic energy density in both simulations are also very similar.

Three stages of magnetic energy growth are also supported by the time evolution of magnetic spectrum.
 Figure  \ref{fig:sp} shows how magnetic spectrum $E_b(k)$ of Run 512-$B_{max}$0.01 evolves with time. 
When $t \lesssim 10$, stretching happens near the dissipation scale and, therefore, 
magnetic energy spectrum peaks there.
As magnetic field is amplified by stretching, the magnetic spectrum moves upward without changing its shape much
(compare, for example, spectra for t=0.24 and t=2.4).
When magnetic spectrum `touches' the velocity spectrum $E_v(k)$ at the dissipation scale,
the exponential growth ends and the slower linear growth stage follows.
Note that, when
magnetic spectrum `touches' the velocity spectrum at the dissipation scale,
the magnetic energy density is approximately $\sim (k_d/k_L)^{2/3}$ times the kinetic one, where
$k_d$ is the dissipation scale wavenumber and $k_L\sim 2.5$, 
and it becomes `visible' in Figure \ref{fig:comp_512}.
The long dashed curves
are the spectra of velocity and magnetic field during the saturation stage.
Velocity spectrum $E_v(k)$ peaks at the energy injection
scale and 
 magnetic spectrum $E_b(k)$ peaks at a smaller scale.
The logarithmic slope of the velocity spectrum during the saturation stage is steeper than $-5/3$ for
$k\lesssim 10$ and gets shallower for $k\gtrsim 10$.
The behavior of magnetic energy spectrum is very similar to that of 
a uniform seed field case (see, for example,
Cho et al.~2009).

\begin{figure*}
\includegraphics[width=0.95\textwidth,bb=0 0 1050 350]{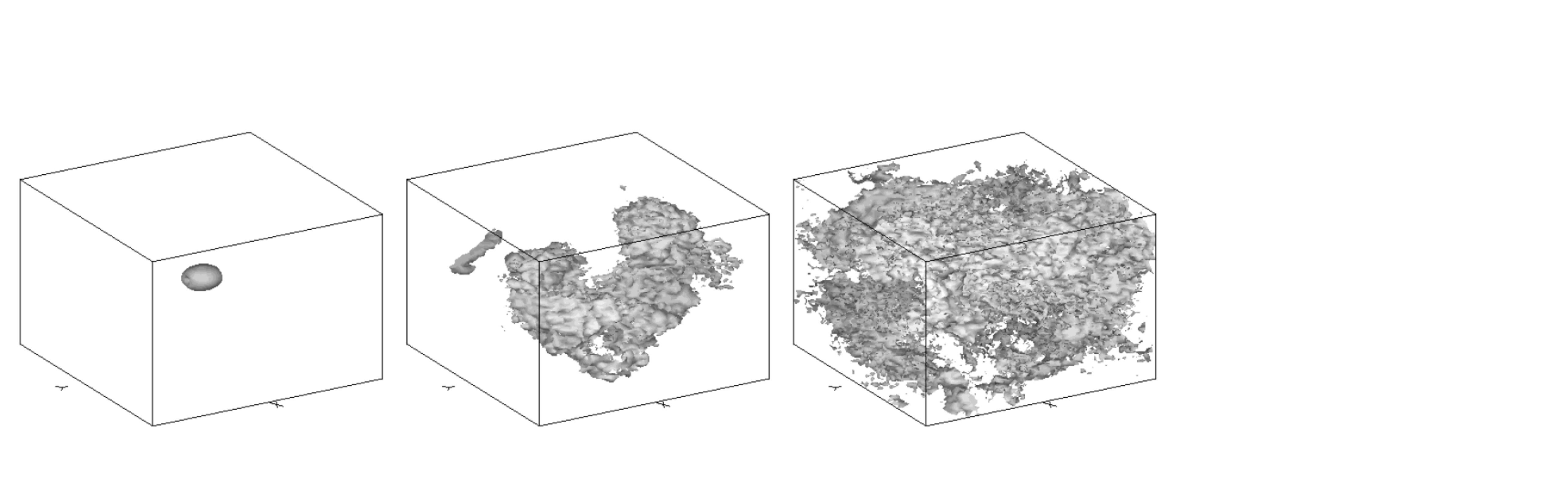}
\caption{Turbulence diffusion of magnetic field.  From left to right panels, snapshots at $t$=0, 1.2, and 2.4 are
   shown. The seed magnetic field has a doughnut shape at t=0.
    At $t=2.4$, the localized seed field fills the most of the computational box.
    Here time (t) is measured in units of $L/v$.
   {}From 256-$B_{max}$0.01.}
\label{fig:diffusion}
\end{figure*}

\begin{figure}
\includegraphics[width=0.49\textwidth,bb=0 0 500 400]{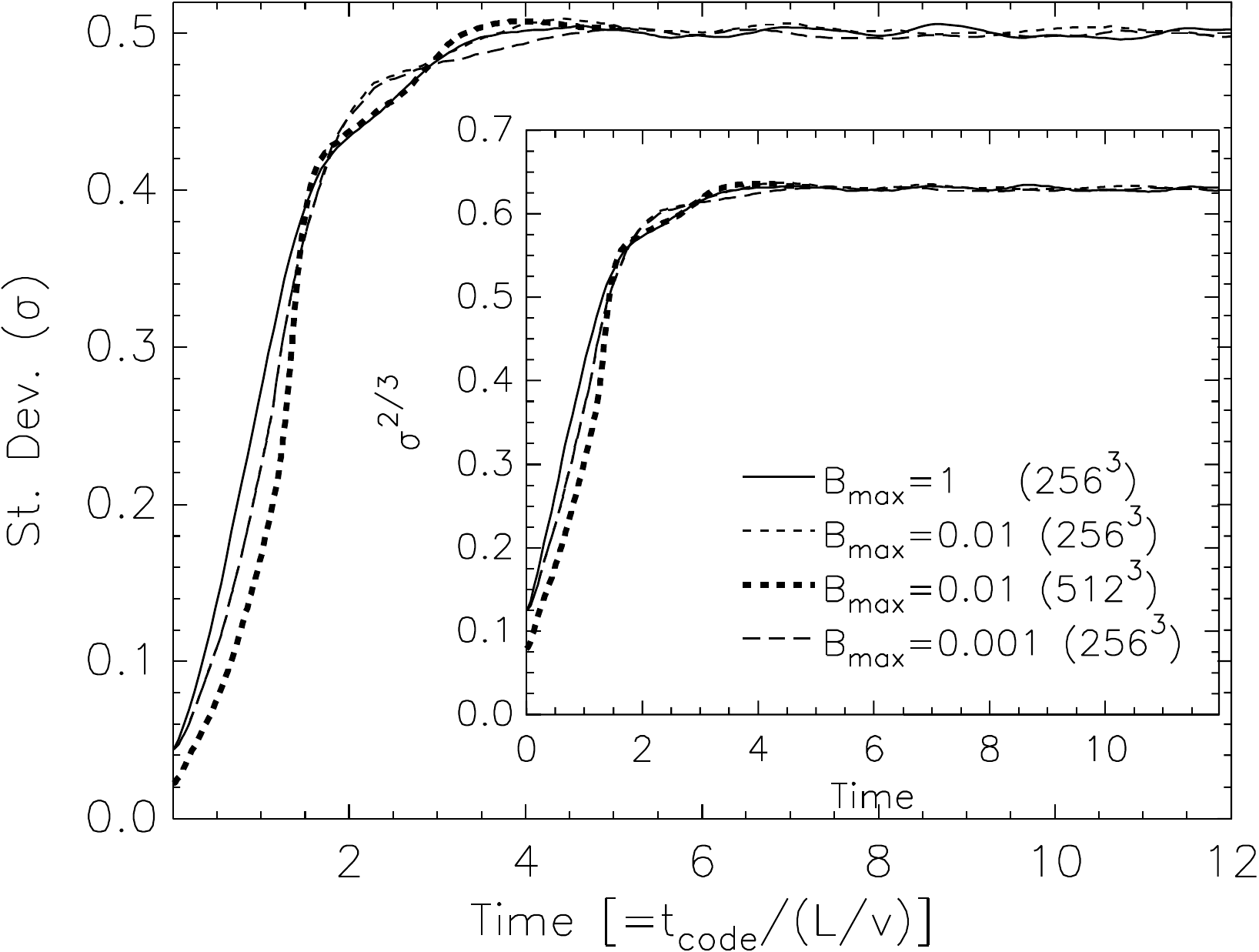}
\caption{Time evolution of the standard deviation of magnetic field distribution.
   The standard deviation is normalized by the box size, so that
   it becomes 0.5 when a homogeneous magnetic field fills the whole box.
   The standard deviation rises quickly when $t\lesssim 1.5$ large-eddy turnover times, 
   which happens when
    the localized magnetic field fills the whole energy-containing eddy.
   After $t\sim 1.4$ large-eddy turnover times, the diffusion rate of the magnetic field slows down. 
   It takes  $\sim 3$ large-eddy turnover times for the magnetic field to fill the whole box.
  Results are from 256-B$_{max}$0.001, 256-B$_{max}$0.01, 256-B$_{max}$1, and 512-B$_{max}$0.01. }
\label{fig:sig}
\end{figure}

So far, we have found surprising similarities 
between localized and uniform seed magnetic fields.  
They both follow similar growth stages,  show similar growth rates at the linear growth stage, 
reach similar levels of magnetic energy saturation, and
exhibit similar time evolution of magnetic spectrum.
Then why are they all similar?
The answer is fast magnetic diffusion.
Turbulence enhances diffusion processes.
Figure \ref{fig:diffusion} visualizes fast magnetic diffusion in Run 256-B$_{max}0.01$.
The localized initial magnetic field (left panel) spreads out by turbulent motions.
Note that it takes only $\sim 2.4$ large-eddy turnover times 
for the localized initial magnetic field to fill virtually the whole box (see the right
panel).
The large-eddy turnover time is defined as $L/v$, where $L$ ($\sim 2.5$) is the outer scale and $v$ 
($\sim 1$) the r.m.s.
velocity.
Therefore, turbulence makes the localized seed magnetic field fill the whole system 
within less than $\sim$3 large-eddy turnover times.
After magnetic field fills the whole box, time evolution of magnetic energy density should be very similar
to that of uniform or spatially homogeneous magnetic field cases (see, for example, Cho et al. 2009).

Figure \ref{fig:sig} confirms fast diffusion of magnetic field, in which we plot
the standard deviation, $\sigma$, of magnetic field distribution:
\begin{eqnarray}
     \sigma = ( \sigma_x^2+\sigma_y^2 + \sigma_z^2 )^{1/2}, \\
    \sigma_i^2 =\frac{ \int ( x_i -\bar{x}_i)^2 |{\bf B}({\bf x},t)|^2 d^3x }{ \int  |{\bf B}({\bf x},t)|^2 d^3x }, \\
  \bar{x}_i=\frac{    \int  x_i  |{\bf B}({\bf x},t)|^2 d^3x }{ \int  |{\bf B}({\bf x},t)|^2 d^3x }, 
\end{eqnarray}
where $i=$x, y, and z and $\sigma$ shown in the plot is normalized by the box size.
Initially the standard deviation rises very quickly, which is due to diffusion of magnetic field within an
energy-containing eddy. The duration of the quick rise is about 1.5 large-eddy turnover times, if we define
the large-eddy turnover time as $L/v$.
The slower growth after $t/(L/v)\sim 1.5$ is due to a slower diffusion of magnetic field 
on scales larger than the
outer scale of turbulence.
The behavior of $\sigma$ is not sensitive to the value of B$_{max}$.

Richardson's law states that the mean square separation of two passive particles, $\sigma_m$, satisfies
\begin{equation}
      \frac{ d \sigma_m }{dt} \sim \epsilon^{1/3} \sigma_m^{1/3}.
\end{equation}
Therefore, 
we may write
\begin{equation}
       \sigma(t)^{2/3}-\sigma(t_0)^{2/3}  \propto  t-t_0.
\end{equation} 
(see, for example, Lesieur 1990; see also Cho et al. 2003).
The inset shows $\sigma^{2/3}$ of magnetic field distribution roughly follows the Richardson's law
 when $t/(L/v) \lesssim 1.5$.

\subsection{Turbulence with a high magnetic Prandtl number ($\nu >> \eta$)}
The viscosity in galaxy clusters may not be negligible 
(Schekochihin et al. 2004; Ruszkowski et al. 2004; Reynolds et al. 2005; Subramanian, Shukurov \& Haugen 2006), 
while the magnetic diffusivity is still very small.
In fact, if we use the Spitzer (1962) formula for the viscosity, the Reynolds number $Lv/\nu$, where 
$L$ is the outer scale of turbulence in a cluster, 
$v$ the r.m.s. velocity, and $\nu$ the viscosity, is less than $\sim 10^3$.
Therefore, the magnetic Prandtl number ($\nu/\eta$) may be very large in the ICM.
In this subsection, we consider growth of localized seed magnetic fields in turbulence with
non-negligible viscosity.
The Reynolds number ($Lv/\nu$) we use in this section is  $\sim 100$.
The magnetic diffusivity is negligibly small because we use a hyper-diffusion.

\begin{figure*}
\begin{tabbing}
\includegraphics[width=0.49\textwidth,bb=0 0 500 360]{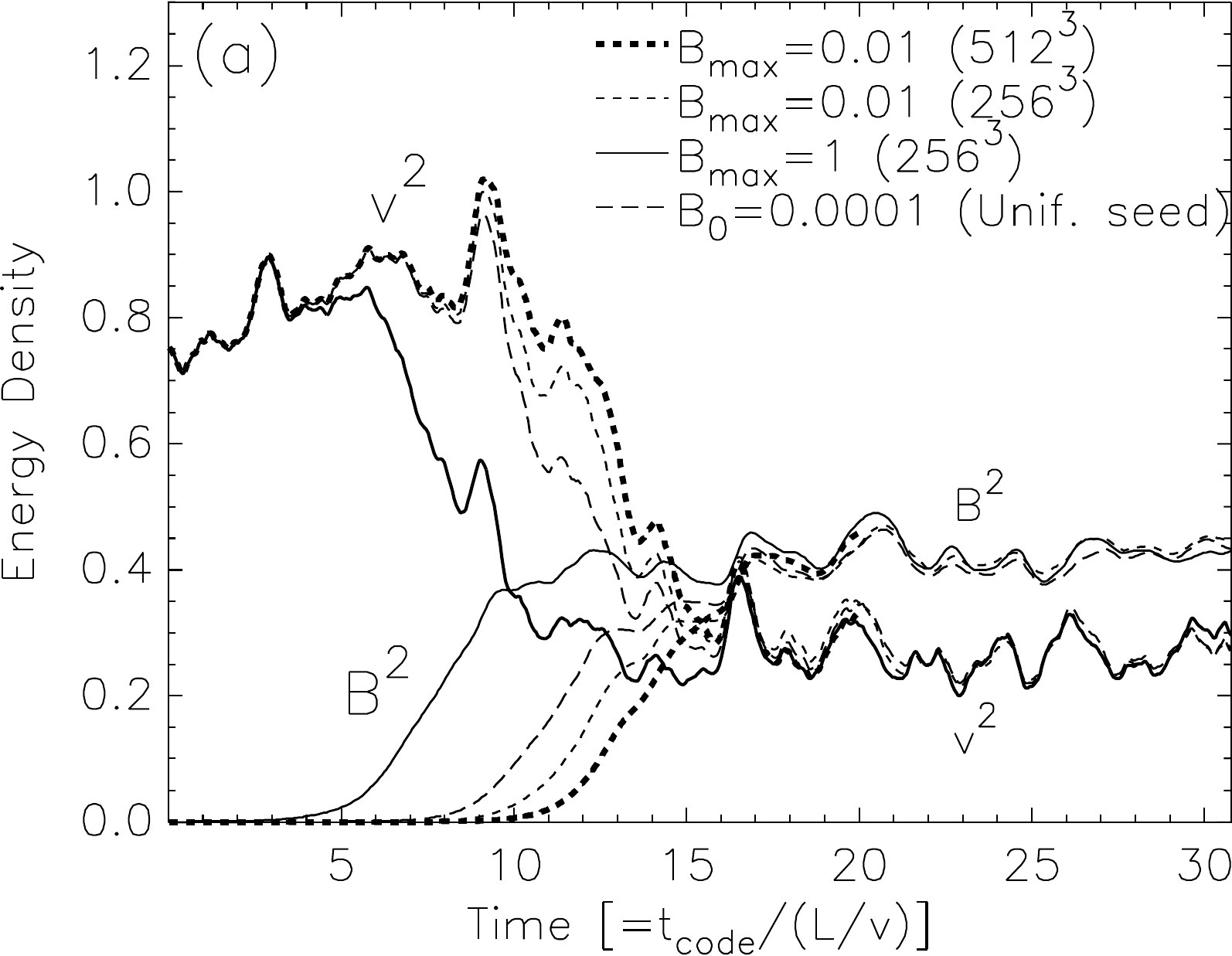}
\=
\includegraphics[width=0.49\textwidth,bb=0 0 500 360]{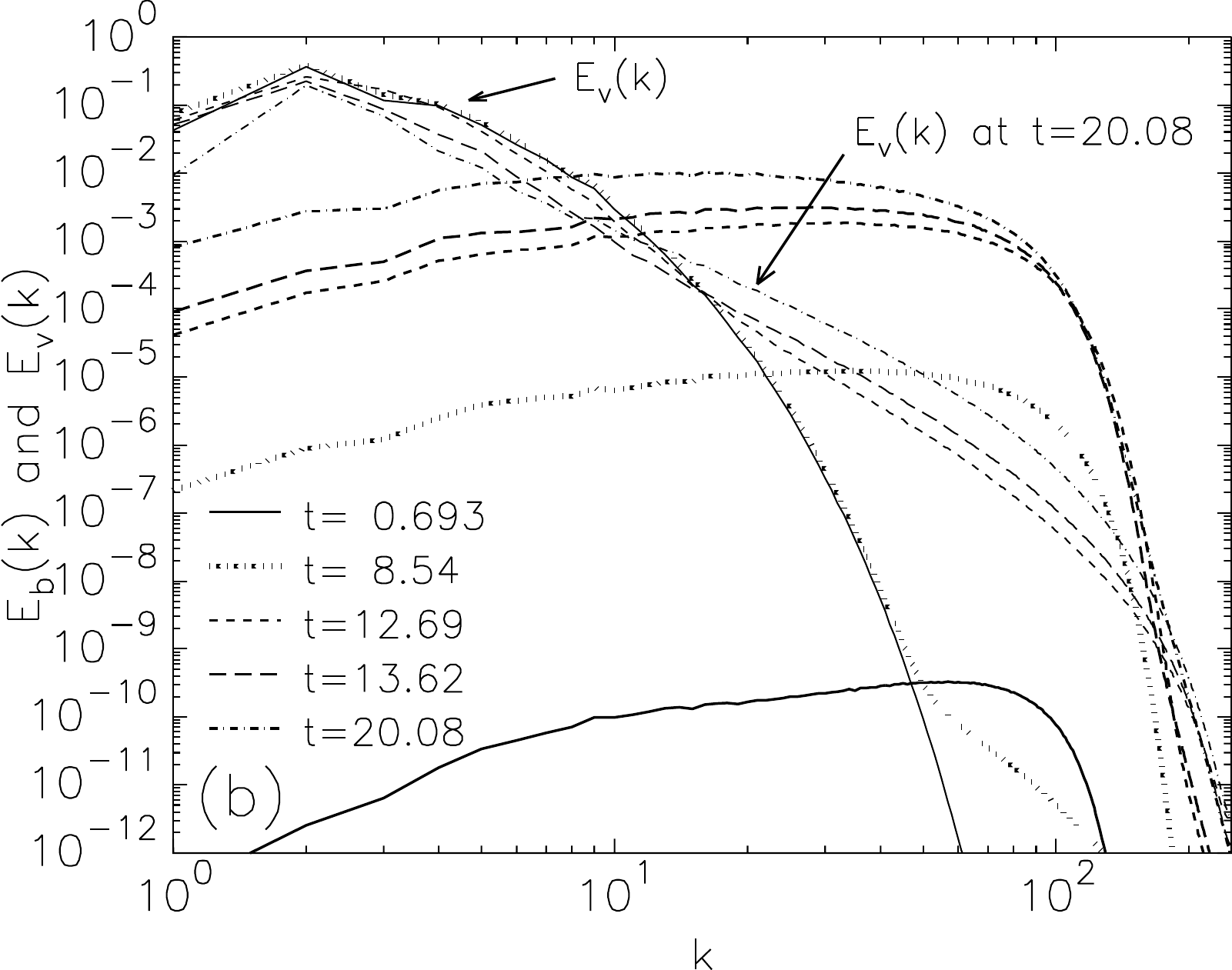} \\
\includegraphics[width=0.49\textwidth,bb=0 0 500 360]{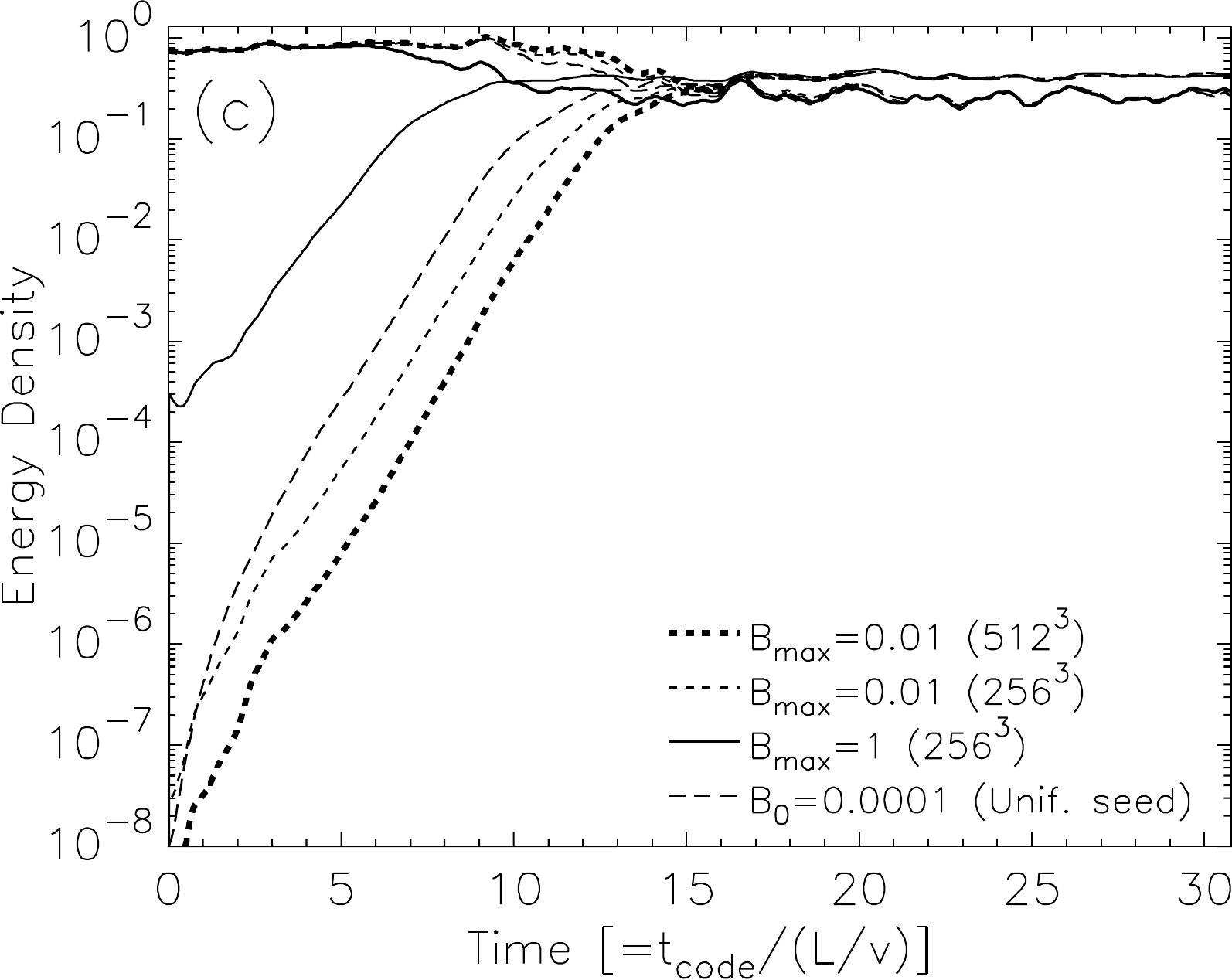} 
\>
\includegraphics[width=0.49\textwidth,bb=0 0 500 360]{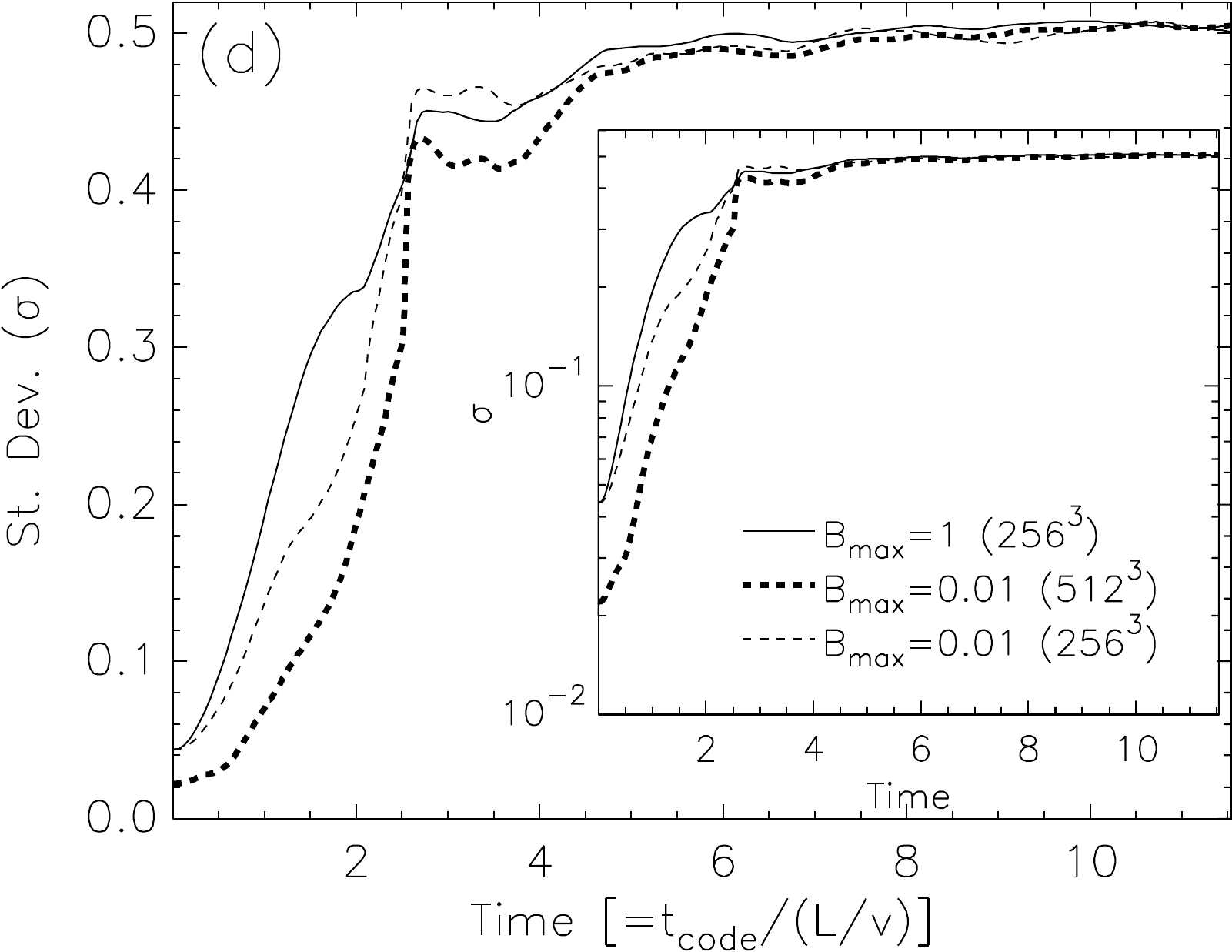} 
\end{tabbing}
\caption{High magnetic Prandtl number turbulence ($\nu \gg \eta$).
               (a) Time evolution of $B^2$ and $v^2$.
               In all simulations, the Reynolds number is $\sim 100$ and the magnetic Reynolds number is very large
              because we use a hyper-diffusivity.
              Except in case of $B_0=0.0001$,  where we use a uniform seed magnetic field,  we use a localized 
             seed magnetic field.
             Time is given in units of the large-eddy turnover time defined by $L/v$, where
     $L$ ($\sim$ 2.5) is the outer-scale of turbulence and
     $v$ ($\sim \sqrt{0.9}$) is the r.m.s. velocity \textit{before} the saturation stage.
     We observe fast growth of magnetic field. 
             Results are from Runs 512-B$_{max}$0.01-Pr, 256-B$_{max}$0.01-Pr, 
                             256-B$_{max}$1-Pr, and REF256-Pr.
     (b) Time evolution of velocity and magnetic spectra.
         Note that the velocity spectrum does not change much during the exponential growth stage.
         {}From Run 512-B$_{max}$0.01-Pr.
     (c) Same as (a), but the vertical axis is drawn in logarithmic scale.
         Note that all simulations with weak $B_{max}$ have similar growth rates during the exponential stage.
         This is because the stretching scales are virtually same in the simulations.
         In Run 512-B$_{max}$0.01-Pr, the exponential growth stage ends at $t\sim$12.
               (d) Standard deviation of magnetic field distribution.
                The localized seed magnetic field fills the whole box within $\sim$3 to 4 large-eddy turnover times.
                }
\label{fig:pm}
\end{figure*}

Figure \ref{fig:pm}(a) shows time evolution of $B^2$ and $v^2$.
Due to non-negligible viscosity in those simulations,
small-scale velocity is strongly damped and
the inertial range of turbulence is poorly resolved.
Indeed the velocity spectrum at t=0.0693 (solid curve) in Figure \ref{fig:pm}(b) confirms this.
Therefore, 
stretching of magnetic field lines happens at a scale very close to
the outer scale of turbulence and
the growth rate of magnetic energy density is mostly exponential (see Figure \ref{fig:pm}(c); see also
next paragraph for further discussions).
Figure \ref{fig:pm}(a) shows that it takes less than $\sim$15 large-eddy turnover times for the runs
to reach the saturation stage.
Comparing with Figure \ref{fig:en_many}, we note that 
growth of magnetic field is at least as fast as that of unit magnetic Prandtl number cases.
Note that the large-eddy turnover time $L/v$ in high magnetic Prandtl number cases is approximately 2.6 
in code units 
before saturation and  4.5 in code units after saturation.

It seems that growth of a localized seed magnetic field 
also happens in 3 stages in a high magnetic Prandtl number turbulence\footnote{
   For a related discussion on the case of a weak uniform seed magnetic field, 
   see, for example, Schekochihin \& Cowley (2007).}.   
\begin{enumerate}
\item{First, we can clearly observe the exponential growth stage in Figure \ref{fig:pm}(c). 
    The exponential growth stage ends when the magnetic energy density becomes comparable to
    the kinetic energy density at the stretching scale.
    Since the stretching scale is very close to the outer scale of turbulence
    in our simulations,
    the exponential stage ends when magnetic energy density becomes comparable to the 
    \textit{total} kinetic energy density.
    Therefore, the exponential stage  
    takes most of the growth time in the high magnetic Prandtl number turbulence.
    Velocity spectrum does not change much during the exponential stage
    (compare $E_v(k)$'s at t=0.693 and 8.54 in Figure \ref{fig:pm}(b)), which
   means that  magnetic back-reaction is negligible  during the exponential growth stage.}
\item{Second, 
    after the exponential growth stage, a slower growth stage follows.
    In Run 512-B$_{max}$0.01-Pr, the slower growth stage is relatively short
    due to proximity between the outer scale and the initial stretching scale:
    it begins at t$\sim$12 and ends at t$\sim$17.
    According to Figure \ref{fig:pm}(a), the growth rate during the slower growth stage is roughly linear.
    The growth rate of $B^2$ during the slower growth stage
    can be estimated as follows.
    If stretching happens at a scale $ l $, we can write
    \begin{equation}
      \frac{ d B^2 }{ dt } \propto \frac{ B^2 }{ l /v_l } = \frac{ B^2 v_l }{ l }  \label{eq11}
    \end{equation}
    (see Schekochihin \& Cowley 2007),
    where $v_l$ is the r.m.s. velocity at the scale $l$.
    The stretching scale $l$ increases during the slower growth stage.
    The peak of the magnetic spectrum in Figure \ref{fig:pm}(b) 
    moves to smaller wavenumbers (compare magnetic spectra at t=12.69, 13.62, and 20.08), 
    which might indicate increase of the stretching scale. 
    The velocity spectrum between $k\sim$3 and $k\sim$10 becomes quickly suppressed
    as soon as the slower growth stage begins
    (compare velocity spectra at t=12.69 and t=13.62).
    The wavenumbers at which velocity spectrum is suppressed might have something to do with
    the stretching scale during the exponential growth stage.
    After the suppression, the stretching scale is clearly the outer scale of turbulence (see the shape of
    $E_v(k)$'s at t=13.62 and 20.08).
    Therefore, the behavior of velocity spectrum also suggests increase of $l$.
    We also note that decrease of $v$ is not negligible
    during the slower growth stage:
    $v^2$ drops fast after $t\sim$12 in Run 512-B$_{max}$0.01-Pr.
    All in all, increase of $B^2$, decrease $v$ ($\sim v_l$), and increase of $l$ make
    the right-hand side of Equation (\ref{eq11}) roughly constant, which results in a roughly linear
    growth rate of $B^2$ in our simulations.}
    
\item{Third, 
    after the slower growth stage, the saturation stage begins.
    In Run 512-B$_{max}$0.01-Pr, it begins at t$\sim$17.}
\end{enumerate}

During the exponential growth stage,
we can write $B^2 \sim B^2(t=0) \exp(t/\tau)$,
where $\tau$ is approximately a constant times the large-eddy turnover time\footnote{
    In our simulations, $\tau$ might be the eddy turnover time at $k\sim$10 (see the previous footnote)}.
Therefore, the growth time of magnetic field depends on the value of $B_{max}$.
Indeed, the figure shows that the run with $B_{max}=1$ reaches 
the saturation stage more quickly than smaller $B_{max}$ cases.
Run 512-B$_{max}$0.01-Pr (thick dashed curves) reaches the saturation stage later than
   Run 256-B$_{max}$0.01-Pr (thin dashed curves), because the size of the initially magnetized region 
   normalized by the system size is smaller in the former, which makes 
   $B^2(t=0)$ smaller in the former.
   If the seed magnetic field is extremely weak,
   it is possible that the growth of magnetic field is slower in high magnetic Prandtl number
   turbulence than in the unit Prandtl number cases with the same seed magnetic field.
   
As in the unit Prandtl number cases, the seed magnetic fields fills the computational box
very fast. According to Figure \ref{fig:pm}(d), 
it takes less than$\sim 4$ large-eddy turnover times
for the seed fields to fill the whole box.
Roughly speaking, the standard deviation of magnetic field distribution, $\sigma$,
grows exponentially when $B_{max}$ is weak.
Consider a magnetized region of size $\sigma$ with $\sigma<L$.
Then, since stretching happens near the outer-scale of turbulence, 
the timescale for a substantial change of $\sigma$ 
is roughly the large-eddy turnover time $L/v$.
Therefore, 
we have the relation 
\begin{equation}
    \frac{\Delta \sigma }{ \Delta t} \propto \frac{\sigma}{L/v} \propto \sigma,
\end{equation}
which implies exponential growth of $\sigma$ when $\sigma < L$.
The inset of Figure \ref{fig:pm}(d), in which the vertical axis is drawn in logarithmic scale, supports
exponential growth of $\sigma$.

\begin{figure*}
\includegraphics[width=0.49\textwidth,bb=0 0 500 400]{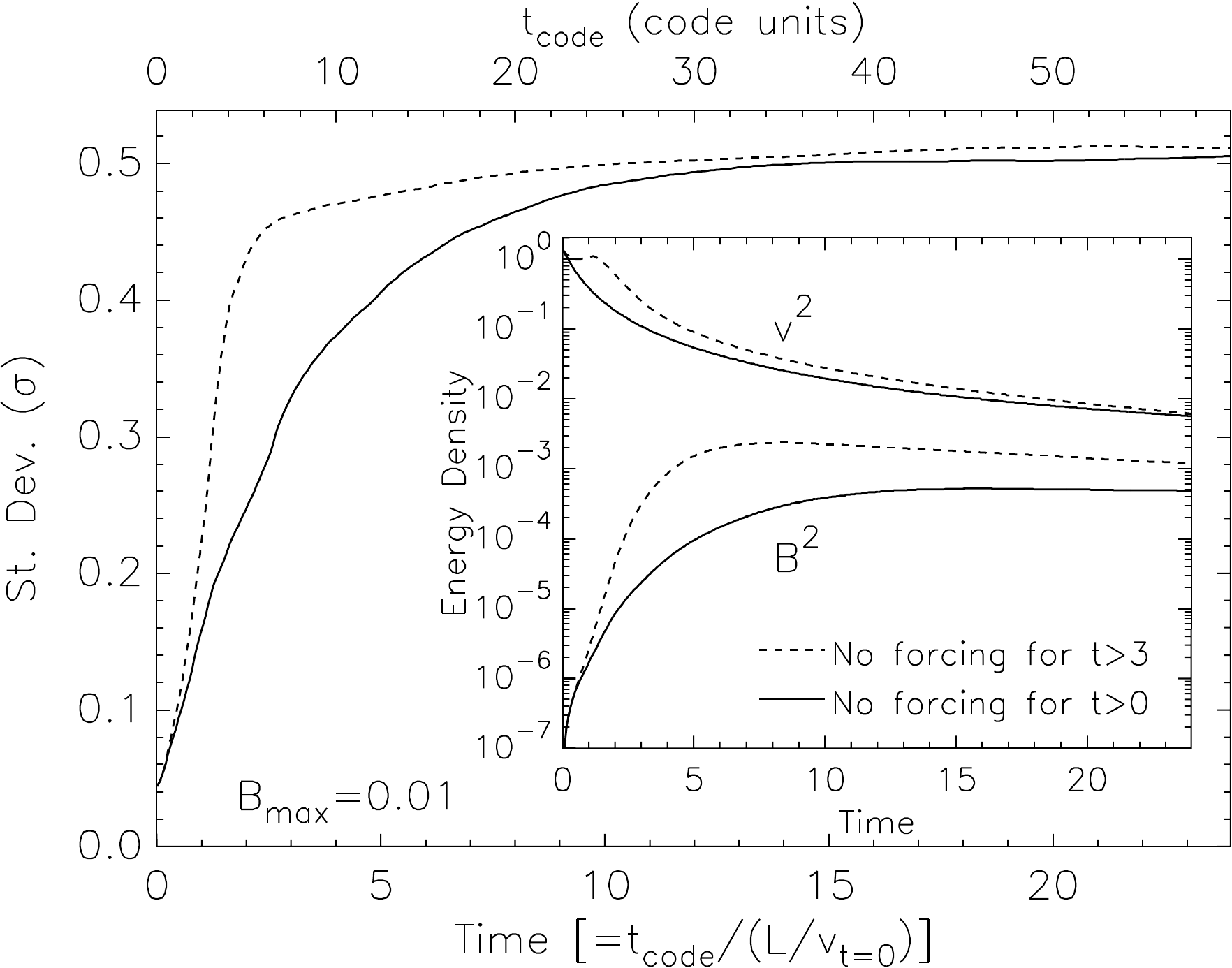} 
\hfill
\includegraphics[width=0.49\textwidth,bb=0 0 500 400]{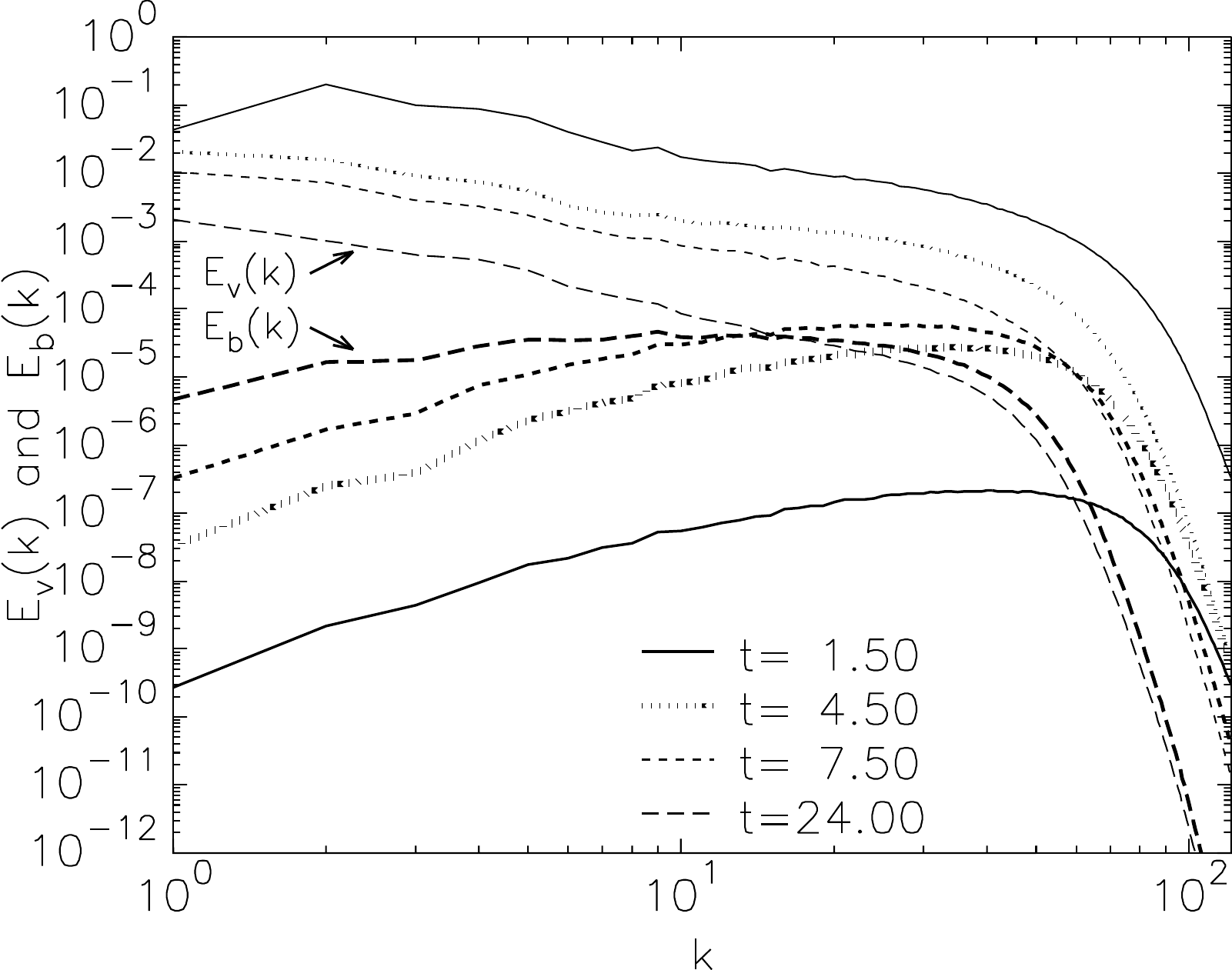} 
\caption{Purely decaying turbulence (Run 256-B$_{max}$0.01-f0; solid curves) and
               turbulence with short-lasting driving (Run 256-B$_{max}$0.01-f3; dashed curves).
               In the former, we let the turbulence decay at t=0.
               In the latter, we drive turbulence for $ t \leq 3$ and we stop driving after $t=3$.
              (Left) Time evolution of energy densities (actually, $B^2$ and $v^2$) 
              and the standard deviation of magnetic field distribution.
              Note that magnetic field spreads and amplifies even in purely decaying turbulence.
              (Right) Time evolution of energy spectra (from Run 256-B$_{max}$0.01-f3).
              Time is given in units of $L/v_{t=0}$ ($\sim$ 2.5 in code units), where $v_{t=0}$ ($\sim$ 1) is the  
              velocity at t=0.
}
\label{fig:decay}
\end{figure*}
\begin{figure*}
\includegraphics[width=0.49\textwidth,bb=0 0 500 400]{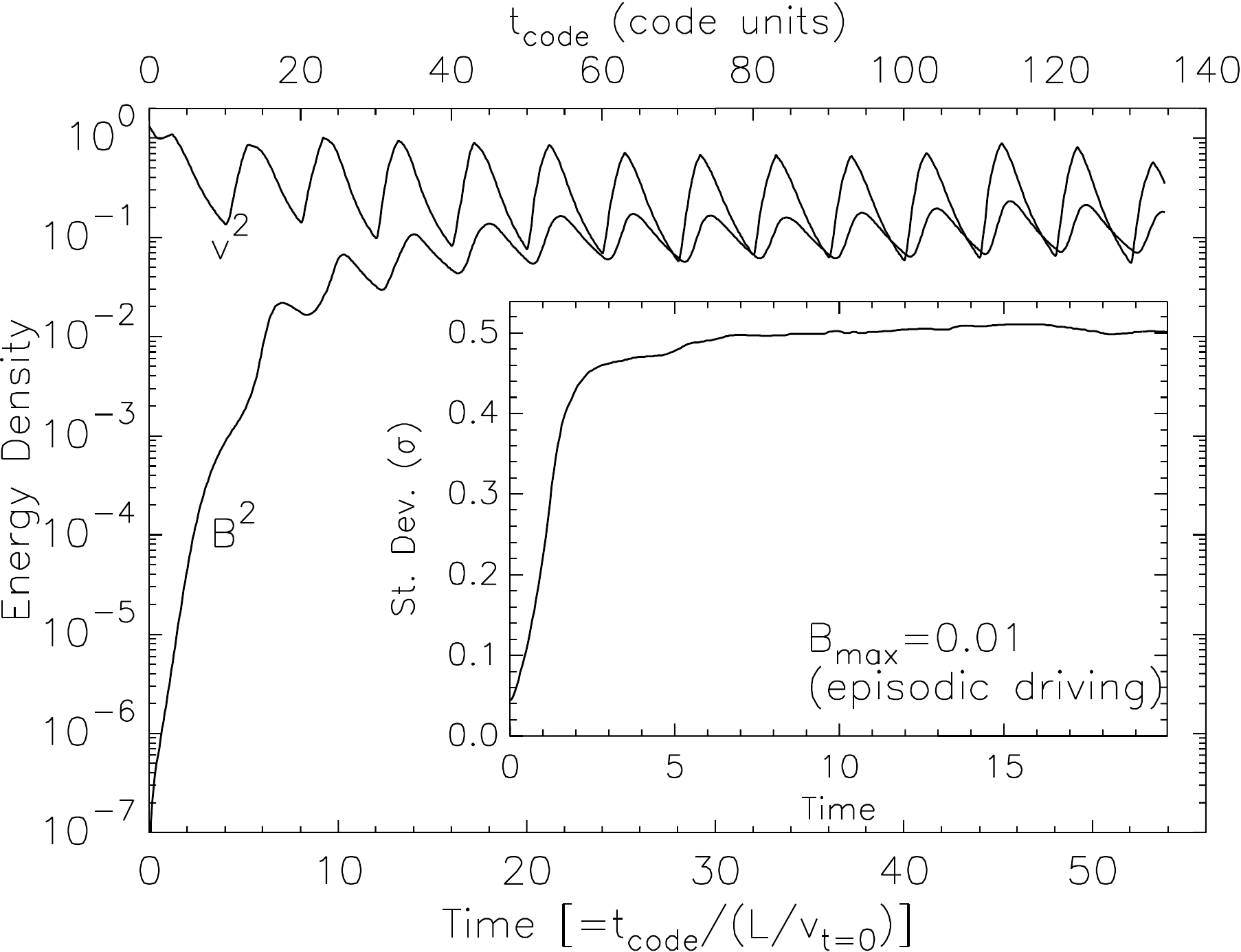} 
\hfill
\includegraphics[width=0.49\textwidth,bb=0 0 500 400]{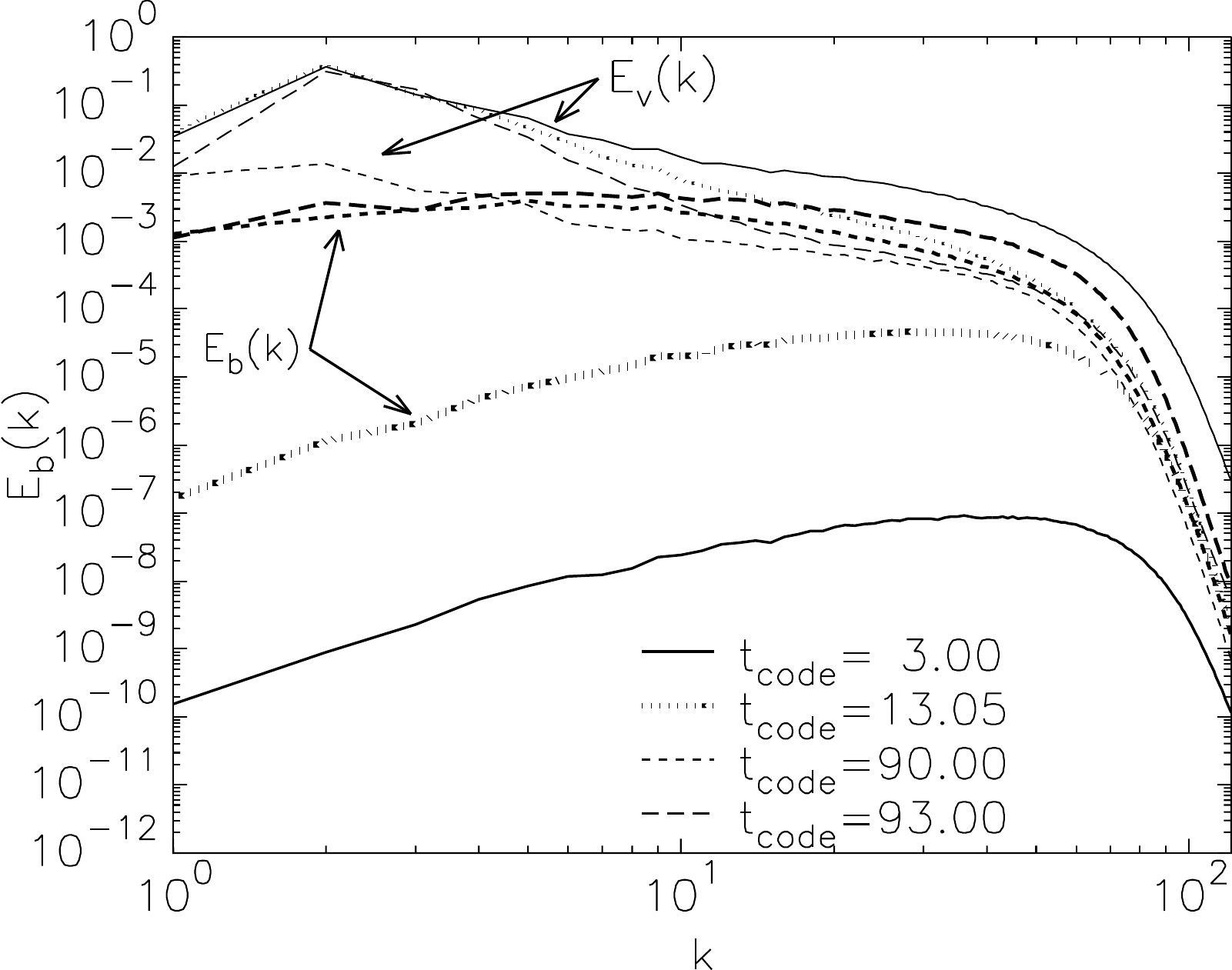} 
\caption{
      Turbulence with episodic forcing (Run 256-B$_{max}$0.01-f$_{epi}$).
      We repeat forcing (3 code time units) and no-forcing (7 code time units) every 10 code time units. 
      (Left) Time evolution of energy densities (actually, $B^2$ and $v^2$)
       and the standard deviation of magnetic field distribution.
       We observe efficient growth and diffusion of magnetic field.
       (Right) Time evolution of energy spectra.
}
\label{fig:epi}
\end{figure*}

\subsection{Decaying turbulence}

If the ICM turbulence is driven by temporarily intermittent events, such as mergers,  the ICM will undergo
disruption and relaxation repeatedly.
In this subsection, we study growth of a localized seed magnetic field in turbulence with episodic driving.

To see if a single merger can disperse a localized seed magnetic field, we first perform
a decaying turbulence simulation (Run 256-B$_{max}$0.01-f0). 
At t=0, we have a fully developed turbulence and a localized seed magnetic 
 field with B$_{max}$=0.01.
The shape of the seed magnetic field  is given in Equation (\ref{eq:b_shape}).
Then we let the turbulence decay without forcing it.
The solid curve in the left panel of Figure \ref{fig:decay} shows that the standard deviation $\sigma$
of the magnetic field distribution gradually rises and converges to the value of 0.5 
within $\sim$10 \textit{initial} large-eddy turnover times ($L/v_{t=0}$).
Therefore, we can conclude that even a decaying turbulence can make a localized seed field 
fill the whole system if
the driving scale of turbulence is comparable to the system size.
The inset in Figure \ref{fig:decay} shows that
$v^2$ immediately decreases as turbulence decays, while $B^2$
rises for $\sim 10$ large-eddy turnover times and then levels off.
As a result, the ratio of magnetic to kinetic energy densities increases.

The results of decaying turbulence simulation imply that a single merger (with magnetic field injection
at the end of the merger) would be enough to 
magnetize the whole ICM.
Consider a localized seed magnetic field in a 
cluster with driving scale of $\sim$400 kpc and initial velocity dispersion of $\sim 400$ km/sec.
Then the large-eddy turnover time is $\sim 10^9$ years.
Our result in Figure \ref{fig:decay} suggests that the decaying turbulence can 
magnetize surrounding $\sim$1 Mpc region of the cluster within the age of the universe. 

Of course, a faster magnetization is possible when we apply additional forcing.
For example, we may allow for short-lasting driving after we inject a localized seed magnetic field, which
mimics a seed magnetic field injection \textit{during} a merger.
We perform a simulation to see the effect of additional forcing (Run 256-B$_{max}$0.01-f3).
The numerical setups are identical to those of the decaying turbulence simulation.
However, in this case we drive turbulence for 3 code time units, 
which is slightly larger than one large-eddy turnover time $L/v \sim 2.5$ code time units.
After $t=3$ code time units, we turn off forcing and  let the turbulence decay.
We plot the result in the left panel of Figure \ref{fig:decay} (see the dashed curve), 
where we can see that $\sigma$ rises to a value very close
to the saturation value of 0.5 within a couple of large-eddy turnover times. 
Note that the strength of turbulence is not a crucial factor here.
Whatever the strength of the turbulence is, a decaying turbulence can virtually seed the whole system within 
a couple of large-eddy turnover times.
But, if the strength of turbulence is very weak, the large-eddy turnover time can be very long.
The inset shows that magnetic energy density rises for $\sim 10$ code time units and then levels off, while 
kinetic energy density begins to decrease right after we turn off forcing at $t=3$ (in code units).
The magnetic energy density in this simulation (dashed curve) is slightly higher than
that of purely decaying turbulence (solid curve).
{}From this simulation and the purely decaying turbulence simulation,
we can conclude that even a single merger is marginally enough to seed
 the whole cluster.  
However, the resulting magnetic field in this case may not be so strong (see values of $B^2$ in the inset).

The right panel of Figure \ref{fig:decay} shows time evolution of magnetic spectrum in 
Run 256-B$_{max}$0.01-f3, which corresponds to the dashed curves in the left panel\footnote{
   The time evolution of magnetic spectra in Run 256-B$_{max}$0.01-f0 is qualitatively similar.
   The reason we show spectra of Run 256-B$_{max}$0.01-f3 is that
   they evolve more quickly due to the additional forcing.
}.
When $t \lesssim 7.5 L/v_{t=0}$, 
magnetic energy spectrum peaks near the dissipation-scale wavenumber, which implies
stretching is most active there, and
the magnetic spectrum moves upward without changing its shape much.
This behavior of magnetic spectrum is very similar to that of a driven turbulence.
However, there are also differences\footnote{
   Since magnetic field fills most of the computational box reasonably fast (see the left panel)
   and  evolution of the system after magnetic field fills 
   the box is very similar to that of
   a weak uniform seed field case, the discussion below should be also applicable to the
   case with a weak uniform seed field.}.
First, the dissipation scale in the decaying turbulence gradually increases as turbulence decays.
In incompressible hydrodynamic turbulence, the dissipation scale is $\sim (Lv/\nu)^{3/4}$ times
smaller than $L$.
Therefore, as $v$ goes down, the dissipation scale goes up.
The eddy turnover time at the dissipation scale increases as $v$ decreases and the dissipation scale increases.
Second, as the eddy turnover time at the stretching scale (i.e.~at the dissipation scale in this case)
 changes, 
the growth rate of magnetic energy density deviates from an exponential one,
which is clearly observed in the left panel of Figure \ref{fig:decay}.
At $t\sim 7.5L/v_{t=0}$, the magnetic spectrum `touches' the velocity spectrum  
at the dissipation scale. %

After $t\sim 7.5L/v_{t=0}$, the evolution of magnetic spectrum is somewhat complicated.
When we compare magnetic spectra at t=7.5 (dashed curves) and 24 (long-dashed curve), we find that
magnetic spectrum for large $k$ values goes down and that for small $k$ values goes up
 after $t\sim 7.5L/v_{t=0}$.
The spectrum for large $k$ values  decreases as the  velocity spectrum decreases due to decay of turbulence.
On the other hand, the spectrum for small $k$ values increases because of stretching at large scales.
As a result, the ratio $B^2/v^2$ increases and the peak of the magnetic spectrum moves to larger scales.
This behavior of $E_b(k)$ will ultimately stop 
when the ratio $B^2/v^2$ becomes large enough so that stretching 
at the outer scale of turbulence becomes suppressed.
Note that, since magnetic spectrum for small $k$ values increases and that for large $k$ values decreases,
the total magnetic energy density changes more slowly than the kinetic energy density
(see the dashed curves for $t\gtrsim$7.5 in the left panel).

\subsection{Turbulence with episodic driving}

In order to have strong magnetization, we need either a continuous or a frequent episodic forcing. 
The left panel of Figure \ref{fig:epi} shows that,
if there are repeated mergers, we can have a relatively strong magnetic field.
As in other simulations, we have a fully developed turbulence and a localized seed magnetic field at t=0.
In the simulation, we drive turbulence for 3 code time units and let it decay for 7 code time units.
Then we repeat the same thing periodically.
{}From the figure, it is evident that a `saturation stage' is reached at $t\sim 50$ in code units.
After $t\sim 50$ code time units, $v^2$ and $B^2$ do not show a systematic decrease or increase and
  just fluctuate around their average values.
The inset shows that the standard deviation at t=10 (in code units)  is slightly smaller than the saturation value of 0.5.
But the second episode of driving starting at t=10 (in code units) quickly boosts it to the saturation value.
The overall behavior of $B^2$ and $v^2$ is very similar to that in continuous-driving cases.
Note that, during the decay stages after $t\sim 50$ (in code units), magnetic energy density sometimes exceeds
kinetic energy density, which is not observed in continuously driven turbulence (with $\nu=\eta$).
Therefore, it may be used to distinguish the driving mechanisms in the ICM.

The right panel of Figure \ref{fig:epi} shows time evolution of energy spectra.
During the growth stages (i.e. $t\lesssim$50 in code units), the overall 
behavior of magnetic spectrum is very similar to 
that in continuously driven cases.
For example, during the rapid growth stage ($t\lesssim$15 in code units),
magnetic spectrum peaks near the dissipation scale and moves upward without changing the shape much
(see solid and dotted curves that represent $E_b(k)$).
This behavior of magnetic spectrum is very similar to that during the exponential growth stage
in a continuously driven turbulence.
The peak of magnetic spectrum moves to smaller wavenumbers during the slower growth stage 
(15$\lesssim t\lesssim$50), which is similar to the behavior of magnetic spectrum during the `linear' growth stage 
in a continuously driven turbulence.
However, after $t\gtrsim$50, 
magnetic and velocity spectra during the driving episodes are clearly different from 
those during the decay (i.e.~no-forcing) phases.
Let us compare spectra at $t=$90 (dashed curves) and 93 (long-dashed curves).
Since an episode of driving starts at t=90 and ends in t=93 in code units,
the dashed curves are spectra during the decay phase and
the long-dashed curves are spectra at the end of the driving episode.
 Velocity spectrum during the decay phase (dashed curve) is clearly lower than
 that during the driving episode (long-dashed curve).
 Magnetic spectrum during the decay phase is also lower than that during the driving episode.

\section{Discussion}
In this paper, we have found that growth of a localized seed magnetic field is as fast as 
that of a uniform magnetic field.
This result implies that magnetic field ejected from astrophysical bodies
can be a viable origin of magnetic field in the large-scale structure of the universe.

Assuming a uniform seed magnetic field (and turbulence generated by cosmological shocks),
Ryu et al. (2008) showed that turbulence in the ICM has almost reached the saturation stage.
Our current simulations imply that we may arrive at a similar conclusion also in the case of 
a localized seed magnetic field.
Therefore, we expect a strong magnetization in clusters regardless of the shape of the seed field, which makes
the findings in Ryu et al.~(2008) more reliable in clusters.

Since both a uniform and a localized seed fields give similar distributions and strengths 
of magnetic fields in clusters, 
 it may be difficult to tell which seed field is more important.
One may think that  it would be more advantageous to look at 
magnetic field distributions in filaments.
Indeed, we expect that large-eddy motions are slower in filaments and
magnetic fields ejected from astrophysical bodies may not have enough time to fill the whole system
(see also discussions in Br\"uggen et al 2005).
Therefore, magnetic field distribution in filaments may be spatially inhomogeneous  
when seed magnetic fields are ejected from galaxies.
Indeed, if $t/(L/v) \lesssim 2$, we expect to see spatially inhomogeneous
distribution of magnetic fields in filaments (see Figures \ref{fig:diffusion} and \ref{fig:sig};
see also the right panel of Figure \ref{fig:pm}). 
Even if $t/(L/v)$ is larger, driving scale of turbulence can also affect spatial inhomogeneity.
If turbulence in filaments is driven on scales much smaller than the system size of a few Mpc,
magnetic fields may not have enough time to diffuse over un-correlated eddies and, as a result,
we may have spatially inhomogeneous magnetic field distributions.

On the other hand, if turbulence in a  filament is driven by cosmological shocks, 
where driving scale
of turbulence is comparable to the system size, we will have more or less homogeneous 
magnetic field distribution.
Cho \& Ryu (2009) argued that the driving scale of turbulence is comparable to the size of the system 
and that $t/(L/v) \sim 4$ in a typical filament.\footnote{
   Cho \& Ryu (2009) used a different definition of the large-eddy turnover time.
  The large-eddy turnover time $t_{eddy}$ in Cho \& Ryu (2009) is roughly 2.5-3 times
  smaller than our large-eddy turnover time.
  Cho \& Ryu (2009) found that $t/t_{eddy} \sim 10$ for a typical filament, which is equivalent
  to $\sim$4 large-eddy turnover times in our definition here.}
Figure \ref{fig:sig} suggests that 4 large-eddy turnover times are more than enough to magnetize 
the whole outer-scale eddy and also adjacent outer-scale eddies, 
the total size of which is larger than a few Mpc.
Therefore, unless the astrophysical sources that provide seed magnetic fields are very sparse,
magnetic fields fill substantial volume fraction in filaments.

Our results show that magnetic diffusion is fast for B$_{max} \leq 1$.
Note that, when B$_{max}=1$, local magnetic energy density is comparable to kinetic energy density
and magnetic backreaction is not negligible in the local region.
Nevertheless, our simulations show overall magnetic diffusion
by turbulence is not suppressed by the backreaction.
In this paper, we do not consider stronger initial seed magnetic fields because we can guess what will
happen.
In case of B$_{max} > 1$, we expect that magnetic pressure first makes magnetic field spread outward,
which should be faster than turbulence diffusion (see Colbert et al.~1996; Xu et al. ~2010, 2011).
Then, after local magnetic energy density drops below the kinetic energy density, turbulence
becomes the major agent that makes magnetic field spread out, which
we know from our current simulations is fast.
Therefore we expect fast diffusion of the seed magnetic field even cases of B$_{max} > 1$.

We do not assume any specific origin of  turbulence in this paper: unless the 
driving scale is much smaller than the system size, it does not matter much.
Consider a cluster of galaxies. 
If turbulence is initiated by cosmological shocks (Ryu et al.~2003; Pfrommer 2006) or
mergers (De Young 1992; Tribble 1993; Norman \& Bryan 1999; Roettiger, Burns, \& Stone 1999; 
Ricker \& Sarazin 2001), 
we expect that the outer scale of turbulence 
is comparable to the size of the whole system. 
In fact, the outer scale of turbulence observed in some simulations during the formation of galaxy clusters
is up to $\sim$ several 100 kpc (Norman \& Bryan 1999; Ricker \& Sarazin 2001), which is
a few times smaller than the cluster size of $\sim$ Mpc.
In our current simulations, the driving scale is about 2.5 times smaller than the simulation box.
Therefore, our current simulations mimic turbulence generated by cosmological shocks or mergers.
However, it is still possible that turbulence in the ICM is driven by 
infall of small structures (Takizawa 2005),
AGN jets (see, for example, Scannapieco \& Br\"uggen 2008), or
galaxy wakes (Roland 1981; Bregman \& David 1989; Kim 2007).
In this case, the driving scale of turbulence can be much smaller than the size of the whole system.
If we consider a localized seed magnetic field inside an energy-containing eddy, 
we know that the seed field can spread and fill the outer-scale eddy fast.
If the driving scale of turbulence is small, the subsequent stage of magnetic field diffusion
may take a long time to fill the whole system because diffusion of magnetic field over
un-correlated eddies may be slower compared with that inside an energy-containing eddy.
We will address this issue elsewhere (Cho 2012, in preparation).

In this paper, we have considered homogeneous turbulence.
However, turbulence in the ICM is not homogeneous.
The ICM density varies from the core region to the cluster outskirts.
Then how can it affect our results?
The value of the ICM density length-scale may be an important factor\footnote{
In addition, magnetic buoyancy can also be an important factor.
If a localized seed magnetic field is injected into the central region of the ICM, 
magnetic buoyancy will enhance the rate of magnetic diffusion.}.
If the length-scale is larger than the outer scale of turbulence, our results in
this paper will not be affected much.
On the other hand, if the opposite case is true, then it depends on the driving mechanism 
in the following way.
First, if turbulence is initiated by a violent event/events and if the outer scale of turbulence
does not change for a couple of large-eddy turnover times, then
our results may not be affected much.
If the outer scale of turbulence
does not change, a couple of large-eddy turnover times would be enough for magnetic field
to fill the most of the ICM.
Second, if turbulence is initiated less violently, then it may be possible that the effective outer scale
of turbulence 
reduces to a value comparable to the density length-scale.
In this case, effective reduction of the outer scale can make the resulting turbulence 
similar to that with small-scale driving.
Further numerical study will be necessary to test this possibility.

\section{Summary}

In this paper, we have assumed that driving scale of turbulence is comparable to
the size of the system, which can be justified if we consider ICM turbulence driven by
mergers or cosmological shocks.
We have found the following results:

\begin{enumerate}
\item The growth of a localized seed magnetic field is as fast as
          that of a uniform seed magnetic  field.
         
\item Turbulence diffusion of magnetic field is fast.
         When we insert a localized seed magnetic field in a turbulent medium,
          turbulent motions make  it disperse and fill the whole system very fast, within $\sim$3 large-eddy turnover times.
         After magnetic field fills the whole system, the time evolution should be very similar to the case of
         a uniform seed magnetic field. This way, we have a fast growth of the seed magnetic field.
         Our result implies that even in filaments
         the volume filling factor of magnetic fields is of order unity
         if turbulence is driven on scales comparable to the size of the systems.
\item Growth and turbulence diffusion
         of a localized seed magnetic field in a high Prandtl number turbulence is also fast.
\item A localized seed magnetic field can ultimately fill the whole system even in a decaying turbulence.
          Therefore, we may have a wide-spread magnetic field in a cluster that has undergone just
         one major  merger.
         Magnetic field can be injected at any time during the merger: our decaying turbulence
         simulation suggests that magnetic field can be injected even at the end of the merger.
         Although the strength of the resulting magnetic field is very weak in decaying turbulence,
        the wide-spread magnetic field can be used as a seed field in the next episode of merger.
\item Growth of a localized seed magnetic field in case of episodic driving is also fast.

\end{enumerate}

\acknowledgements
This research was supported by National R \& D Program through 
the National Research Foundation of Korea (NRF) 
funded by the Ministry of Education, Science and Technology (No. 2011-0018751).
We thank the anonymous referee for constructive comments.



\begin{deluxetable}{lccccr}
\tabletypesize{\scriptsize}
\tablecaption{Simulations.}
\tablewidth{0pt}
\tablehead{
\colhead{Run} & \colhead{Resolution} & \colhead{$B_{max}$ at t=0\tablenotemark{a}} 
   & \colhead{$B_0$\tablenotemark{b}} &  \colhead{Pr,m}\tablenotemark{c}  & \colhead{forcing} 
}
\startdata
256-B$_{max}$0.001 & $256^3$ & 0.001  & 0. &  1 & continuous \\ 
256-B$_{max}$0.01 & $256^3$ & 0.01  & 0. &  1 & continuous \\ 
256-B$_{max}$0.1 & $256^3$ & 0.1  & 0. &  1 & continuous \\ 
256-B$_{max}$1 & $256^3$ & 1  & 0. &  1 & continuous \\ 
512-B$_{max}$0.01 & $512^3$ & 0.01  & 0. &  1 & continuous \\  \\ 
%
%
256-B$_{max}$0.01-Pr & $256^3$ & 0.01 & 0. &  high & continuous \\ 
256-B$_{max}$1-Pr      & $256^3$ &       1  & 0. &  high & continuous \\ 
512-B$_{max}$0.01-Pr & $512^3$ & 0.01 & 0. &  high & continuous \\ \\ 
256-B$_{max}$0.01-f0 & $256^3$ & 0.01  & $ 0 $ &  1 & no forcing \\ 
256-B$_{max}$0.01-f3 & $256^3$ & 0.01  & $ 0 $ &  1 & $t\leq 3$ in code units \\ 
256-B$_{max}$0.01-f$_{epi}$ & $256^3$ & 0.01  & $ 0 $ &  1 & episodic \\ \\ 
REF256 & $256^3$ & -  & $10^{-4}$ &  1 & continuous   \\  
REF256-Pr  & $256^3$ &    -  & $10^{-4}$ &  high & continuous  
\enddata
\tablenotetext{a}{Maximum value of the random field at t=0.}
\tablenotetext{b}{Strength of the mean field.}
\tablenotetext{c}{Magnetic Prandtl number ($\equiv \nu/\eta$).}
\label{table_1}
\end{deluxetable}

\end{document}